\documentclass[10pt]{article}

\begin{document}

\title{Approaching the event horizon of a black hole}

\author{A.Y. Shiekh\footnote{\tt ashiekh@coloradomesa.edu} \\
              Department of Physics \\ 
              Colorado Mesa University \\
              Grand Junction, CO \\ 
              U.S.A.}

\date{}

\maketitle

\begin{abstract}
It is argued that it takes an infinite amount of external time for a freely falling test particle to reach the event horizon of a classical black hole (which happens in finite faller time), and that in this time the black hole would have evaporated due to Hawking radiation; so the freely falling test particle would itself evaporate at the event horizon, and so not pass through.
\end{abstract}

\section{Introduction}
Many texts on general relativity~\cite{Hartle},~\cite{TaylorWheeler},~\cite{MTW} consider what happens to a freely falling object after it reaches the event horizon of a black hole, which happens in finite faller time. The perspective being taken in this work is that to reach the event horizon takes an infinite amount of external time and that during this time the black hole would have evaporated due to Hawking radiation, and so the freely falling object would itself evaporate at the event horizon, and so not pass through.

The analysis is not original, but the interpretation is believed to be so. Objections such as coordinate singularities and the possible dilemma of the falling object being left below the horizon due to the black hole accreting further matter are tackled.

\section{Approaching the event horizon}
Starting from the Schwarzschild metric in Schwarzschild coordinates and working with geometrized units ($G = 1$, $c = 1$):
\begin{equation}
d s^2 = -\left( 1 - \frac{2M}{r} \right) d t^2 + \left( 1 - \frac{2M}{r} \right)^{-1} d r^2 + r^2 \left(d \theta^2 + \sin^2 \theta \ d \phi^2 \right)
\end{equation}
which for the radial case ($d \theta = 0$, $d \phi = 0$) simplifies to:
\begin{equation}
d s^2 = -\left( 1 - \frac{2M}{r} \right) d t^2 + \left( 1 - \frac{2M}{r} \right)^{-1} d r^2
\label{eqn:metric}
\end{equation}
where $M$ is the mass of the black hole; this metric has coordinate singularities, that may or may not imply a physical singularity.

To show there is divergent time dilation at the event horizon is easiest first done for stationary frames outside a stationary black hole; since we are always above $r = 2M$ any singularity there does not affect the calculation.

\subsection{Stationary frames (shells)}
Consider two stationary ($dr = 0$), and so non-inertial frames, one close to the event horizon at $r_e$ and the other further out at $r_o$; the proper time ($\tau$) for each can be determined from the above metric (equation~\ref{eqn:metric}) using $d\tau^2 = - ds^2$ to yield:
\begin{equation}
- \ d \tau_e^2 = - \left( 1 - \frac{2M}{r_e} \right) d t^2
\end{equation}
and
\begin{equation}
- \ d \tau_o^2 = - \left( 1 - \frac{2M}{r_o} \right) d t^2
\label{eqn:externaltime}
\end{equation}
Now eliminating the coordinate time $t$ yields:
\begin{equation}
d \tau_o = \left( \frac{1 - 2M/r_o}{1 -2M/r_e} \right)^{1/2} d \tau_e
\end{equation}
So one sees a divergence of time dilation, where time when close to the event horizon ($r_e = 2M$) passes divergently slowly relative to time at any position outside the event horizon; one can do the same calculation with Eddington-Finkelstein coordinates to get the same result.

\subsection{Radial Free fall}
The case for light in radial fall is simplest, where $ds^2 = 0$, and equation~\ref{eqn:metric} for the metric leads directly to:
\begin{equation}
\frac{dt}{dr} = - \left( 1 - \frac{2M}{r} \right)^{-1}
\end{equation}
where the negative square root is appropriate for a geodesic moving inward. This integrates (with $t_*$ the integration constant) to yield:
\begin{equation}
t = -r - 2M \ln \left( \frac{r}{2M} -1 \right) + t_*
\end{equation}
which is still in coordinate time $t$; but this can be eliminated for the proper time of a stationary frame at external radius $r_o$ using equation~\ref{eqn:externaltime} above to yield:
\begin{equation}
\tau_o = -\left( 1 - \frac{2M}{r_o} \right)^{1/2} \left(r + 2M \ln \left( \frac{r}{2M} -1 \right) \right) + \tau_*
\end{equation}
The divergent time dilation (in the logarithmic divergence) is seen to be present for light fall and is of the same physical nature as the infinite red shift or infinite escape energy from the horizon, and not due to any physical divergence in the metric, of which there is none at the horizon.

One can do the same analysis for a radially freely falling massive test particle starting from rest at infinity, but since nothing can travel locally faster than the speed of light, a divergence in external time must still be present; quoting the well known result~\cite{Hartle},~\cite{TaylorWheeler},~\cite{MTW}
\begin{equation}
t = 2M \left[ -\frac{2}{3} \left( \frac{r}{2M} \right)^{3/2} - 2 \left( \frac{r}{2M} \right)^{1/2} + \ln \frac{(r/2M)^{1/2} + 1}{(r/2M)^{1/2} - 1} \right] + t_*
\end{equation}
which can be moved to the proper time of a stationary frame at external radius $r_o$ again using equation~\ref{eqn:externaltime} above to yield:
\begin{equation}
\tau_o = 2M \left( 1 - \frac{2M}{r_o} \right)^{1/2} \left[ -\frac{2}{3} \left( \frac{r}{2M} \right)^{3/2} - 2 \left( \frac{r}{2M} \right)^{1/2} + \ln \frac{(r/2M)^{1/2} + 1}{(r/2M)^{1/2} - 1} \right] + \tau_*
\end{equation}
and as expected is seen to be divergent at the event horizon.

\subsection{Generalization}
The source of the logarithmic divergence in coordinate time, which carries over to external time by equation~\ref{eqn:externaltime}, is just the result of the relation~\cite{Hartle},~\cite{TaylorWheeler},~\cite{MTW}
\begin{equation}
\frac{dt}{d\tau} = e \left( 1 - \frac{2M}{r} \right)^{-1}
\end{equation}
for free fall of a massive test particle around a stationary black hole, true for any energy, not just for the particle at rest at infinity; where $e$ is a constant, namely the energy per unit rest mass. This monotonic relationship ensures the ordering of causally connected events is the same for the freely falling and stationary frames.

A similar divergence is picked up for an equatorial orbit around the rotating (Kerr) black hole~\cite{Hartle},~\cite{MTW},~\cite{TaylorWheeler}
\begin{equation}
\frac{dt}{d\tau} = \frac{1}{\Delta} \left[ \left( r^2+a^2+\frac{2Ma^2}{r} \right)e - \frac{2Ma}{r}\ell \right]
\end{equation}
where $a \equiv J/M$ is the angular momentum per unit mass of the black hole, $\Delta \equiv r^2 - 2Mr + a^2$ and $e$ and $\ell$ are constants, the energy and angular momentum per unit rest mass for the falling object.

To quote Kip Thorne~\cite{Thorne} ``When an infinite amount of external time has passed, the particle has experienced only a finite and very small
amount of time. In that finite time, the particle has reached the hole's horizon \ldots''.

\section{The growing black hole, a contradiction?}
The above analysis was for a test particle, namely one that does not result in the growth of the black hole, and one might wonder what happens when the black hole grows due to mass accretion; what becomes of the test particle that then gets left below the outward growing Schwarzschild radius. This does not pose a dilemma because, once a horizon has first formed, then by the above argument, nothing can reach it in finite external time, not even light; as a result the metric there is frozen and unmodified by any in-falling object, so while the Schwarzschild radius grows, the event horizon extends all the way inward from this radius.

Eventually the black hole will stop growing by accretion and start shrinking by evaporation and the faller will then suffer from the same evaporation, having noticed no further time to have gone by; so the time to death happens in finite faller time.

\section{Observation, a distraction?}
While most texts seem to agree that the analysis has the faller approach the event horizon asymptotically in external time, they argue that this is not what would be seen. The external observer would see something different from reality due to the time delay, red-shift and bending of light; while the stationary observer close to the event horizon would also see something different because their time is dilated. All this is true, but does not avoid the issue that in falling to the event horizon infinite external time passes for finite faller time. In the case of special relativity the image of a Lorentz contracted object is seen to be rotated, but one does not then conclude that the contraction is due to the object having actually rotated.

The approach being taken here is to calculate what is actually happening to the freely falling object and not to add the distractions on how the object would be seen to fall.

\section{Conclusion}
A test particle freely falling toward the event horizon of a black hole will fall into a region just outside the event horizon where the time dilation will allow the black hole to age considerably for a short amount of faller time, and during this time the black hole will evaporate due to Hawking radiation and with it so will the faller, it being part of the black hole; as a result it would not get to pass through the horizon.
This is more simply stated as: one cannot pass through a zone of infinite time dilation simply because the zone cannot endure forever and therefore is gone before one is through, and so has not been passed.

The possible problem of the Schwarzschild radius growing and leaving the test particle below the event horizon does not appear as the event horizon is seen to extend all the way below it; so it is argued that the vacuum black hole metric only applies outside the event horizon, and that there is no singularity at $r = 0$.

Einstein and Eddington may have been less troubled by an interpretation of black holes free of singularities such as presented here.

\end{document}